\documentclass{webofc}
\usepackage[varg]{txfonts}   % Web of Conferences font
\usepackage{epsfig}
\usepackage{array}
\usepackage{picture}
\usepackage{subfig}
\usepackage{graphicx}
\usepackage{color}

\newcommand{\beq}{\begin{equation}}
\newcommand{\eeq}{\end{equation}}
\newcommand{\bea}{\begin{eqnarray}}
\newcommand{\eea}{\end{eqnarray}}

\newcommand{\ds}{\displaystyle}

\newcommand{\aasymt}{{\cal A}}

\begin{document}
\title{Correlations and Critical Behavior\\
in Lattice Gluodynamics.}

\author{\firstname{Vitaly} \lastname{Bornyakov}\inst{1,2,3}\fnsep\thanks{\email{bornvit@gmail.com}} \and
        \firstname{Vladimir} \lastname{Goy}\inst{3,4}\fnsep\thanks{\email{vovagoy@gmail.com}} \and
        \firstname{Evgeny} \lastname{Kozlovsky}\inst{1}\fnsep\thanks{\email{Evgeny.Kozlovsky@ihep.ru}}
         \and
        \firstname{Valentin} \lastname{Mitrjushkin}\inst{5}\fnsep\thanks{\email{vmitr@theor.jinr.ru}}
 \and
        \firstname{Roman} \lastname{Rogalyov}\inst{1}\fnsep\thanks{\email{rnr@ihep.ru}}
}

\institute{Institute for High Energy Physics of the NRC ``Kurchatov Institute'', 142281 Protvino, Russia
\and
Institute of Theoretical and Experimental Physics of the NRC 
``Kurchatov Institute'', 117259 Moscow, Russia
\and
Pacific Quantum Center, Far Eastern Federal University, Sukhanova 8,  690950 Vladivostok, Russia
\and
Institut Denis Poisson CNRS/UMR 7013, Universit\'e de Tours, 37200 Tours, France
\and
Joint Institute for Nuclear Research, 141980 Dubna, Russia
          }

\abstract{%
In the Landau-gauge lattice gluodynamics
we find that, both in the SU(2) and SU(3) theory,
a correlation of the Polyakov loop 
with the asymmetry of the $A^2$ gluon condensate as well as 
with the longitudinal propagator makes it possible to determine
the critical behavior of these quantities. 
We discuss finite-volume corrections 
and reveal that they can be reduced 
by the use of regression analysis.  
We also analyze the temperature dependence of low-momenta propagators 
in different Polyakov-loop sectors.
}
\maketitle
\section{Introduction}
\label{intro}

In last decades, the deconfinement transition 
of strong-interacting matter
at finite temperature has received considerable
theoretical and experimental study.
At physical values of the parameters
it is a crossover transition,
which probably coincides with the 
chiral transition.

In the limit of infinitely-heavy quarks,
the fermion degrees of freedom can be neglected and
the crossover transition goes over into the
first-order transition in conventional QCD
and into the second-order transition in 
two-color QCD. For this reason, 
gluodynamics (that is, pure-gauge theory)
is widely employed as a testing tool 
for the studies of the deconfinement transition.

The temperature and volume dependence of the Green's functions 
of gauge fields in the vicinity of the critical temperature $T_c$ is of particular interest.
%have received signifigant attention in the literature \cite{Huber:2018ned}.
%\cite{Alkofer:2006fu,Fischer:2006ub,Huber:2018ned}.
The critical behavior of the gluon and ghost propagators 
was considered, in particular, in \cite{Fischer:2010fx,Maas:2011ez,Aouane:2011fv}.
In Refs.~\cite{Oliveira:2014uga,Silva:2016onh} it was found that, close to criticality, the gluon propagator behaves differently
in different Polykov-loop sectors.

A interesting example of critical behavior is provided by the 
chromoelectric-chromomagnetic asymmetry of the dimension-two
gluon condensate \cite{Chernodub:2008kf}.
Motivation for the studies of the asymmetry and gluon propagators
was also discussed in \cite{Aouane:2011fv,Maas:2011se,Vercauteren:2010rk} 
and references therein.

Recently, it was demonstrated that 
correlations between the Polyakov loop
and the zero-momentum longitudinal propagator 
considerably facilitate the analysis of its critical behavior
both in the SU(2) \cite{Bornyakov:2016geh,Bornyakov:2018mmf}
and SU(3) \cite{Bornyakov:2021pls} gluodynamics.
This made it possible to describe critical behavior
of the asymmetry and the propagator with unprecendented precision.
Here we discuss the assumptions made in 
Refs.\cite{Bornyakov:2016geh,Bornyakov:2018mmf,Bornyakov:2021pls}
and the role of finite-volume effects.

The paper is organized as follows. In the next Section 
we introduce the definition and describe the details 
of our numerical simulations. 
The correlation between the asymmetry ${\cal A}$ and 
the Polyakov loop ${\cal P}$ forms the subject 
of Section~\ref{sec:corr}. 
In Section~\ref{sec:FVE} we consider the finite-volume effects
and discuss how the Polyakov loop determined 
the behavior of the asymmetry and the propagator.
The propagators at nonzero momenta are considered 
in Section~\ref{sec:glpr_pneq0}.
In Conclusions we summarize our findings.

\section{Designations and lattice settings}
\label{sec:def}

We study SU(2) and SU(3) lattice gauge theories with 
the standard Wilson action in the Landau gauge. 

Our calculations are performed on $N_t\times N_s^3$ lattices
($N_t=8$, $N_s=32,48,72,80$ in the SU(2) case 
and $N_t=8$, $N_s = 24$ in the SU(3) case).
The temperature $T$ is given by $~T=1/aN_t~$ where $a$
is the lattice spacing; $L=aN_s$ is the lattice size. 
We use the parameter $\ds \tau = {T-T_c \over T_c}$.

In the SU(3) case we employ the scale fixing procedure 
proposed in \cite{Necco:2001xg} and use the value of the Sommer parameter $r_0=0.5$~fm as in \cite{Bornyakov:2011jm}. 
Making use of $\beta_c=6.06$ and 
$\ds {T_c\over \sqrt{\sigma}}=0.63$,
 Ref.~\cite{Boyd:1996bx} gives $T_c=294$~MeV and 
$\sqrt{\sigma}=0.47$~GeV. We study the range of temperatures 
$0.904 T_c \leq T \leq 1.104 T_c$, the corresponding lattice spacings
$0.076~\mathrm{fm}\leq a\leq 0.093$~fm. 
Some $200$ independent Monte Carlo
gauge-field configurations 
are generated for each of the sectors of the Polyakov loop ${\cal P}$
(also referred to as the center sectors):
\bea\label{eq:sec_def}
(I)   \qquad -\;{\pi\over 3} < &\arg {\cal P}& < {\pi\over 3} \quad\mbox{referred to as}\quad \mathrm{Re}{\cal P}>0  \\ \nonumber
(II)  \qquad\quad    {\pi\over 3} < &\arg {\cal P}& < \pi  \quad\mbox{referred to as}\quad \mathrm{Re}{\cal P}<0 \\ \nonumber
(III) \qquad -\;{\pi}        < &\arg {\cal P}& < -\;{\pi\over 3} \;.  \quad\mbox{referred to as}\quad \mathrm{Re}{\cal P}<0 \nonumber
\eea
We make no difference between sectors $II$ and $III$ becuse the quantities 
under study are independent of $\mathrm{Im} {\cal P}$ as was shown 
in~\cite{Bornyakov:2021pls}.

In the SU(2) case we use the string tension $\sigma=440$~MeV,
which gives~\cite{Bloch:2002we} $T_c=297$~MeV.  We study the range 
$0.9987T_c\leq T \leq 1.0083 T_c$, the corresponding lattice spacings
$a\approx 0.083$~fm. From 400 to 2200 gauge-field configurations 
are simulated for each set of the parameters. In the case of SU(2)
we make no difference between ${\cal P}$ and $\mathrm{Re}{\cal P}$.

Consecutive configurations are separated by $200\div 400$ 
sweeps, each sweep includes one local heatbath update followed
by $N_s/2$ microcanonical updates. 

Definitions of the chromo-electric-magnetic 
asymmetry $\aasymt$ can be found e.g. in Refs.\cite{Chernodub:2008kf,Bornyakov:2016geh}; the definitions of vector potentials $A_\mu(x)$, 
and the description of gauge-fixing procedure 
as well as the expressions for the longitudinal $D_L(p)$
and transverse $D_T(p)$ gluon propagators --- in 
Refs.\cite{Bornyakov:2011jm,Aouane:2011fv,Bornyakov:2021pls}.
Here we only present the expression for the
asymmetry in terms of the propagators
\beq\label{eq:average_bare_asymmetry}
 \aasymt={2(N_c^2-1) N_t \over \beta a^2 N_s^3 }
\left[ 3 (D_L(0) - D_T(0)) + \sum_{p\neq 0}\left(
{3|\vec p|^2 \,-\, p_4^2\over p^2}  D_L(p) - 2  D_T(p)\right)\right] \;.
\eeq

Here we do not consider details of the approach to the continuum limit 
and renormalization considering that the lattices with $N_t=8$ 
(corresponding to spacing $a\simeq 0.08$~fm at $T \sim T_c$) 
are sufficiently fine.

\section{Correlations of the longitudinal propagator and the asymmetry with the Polyakov loop}
\label{sec:corr}

\begin{figure}[h]
\vspace*{-14mm}
\hspace*{-2mm}\includegraphics[width=7cm,clip]{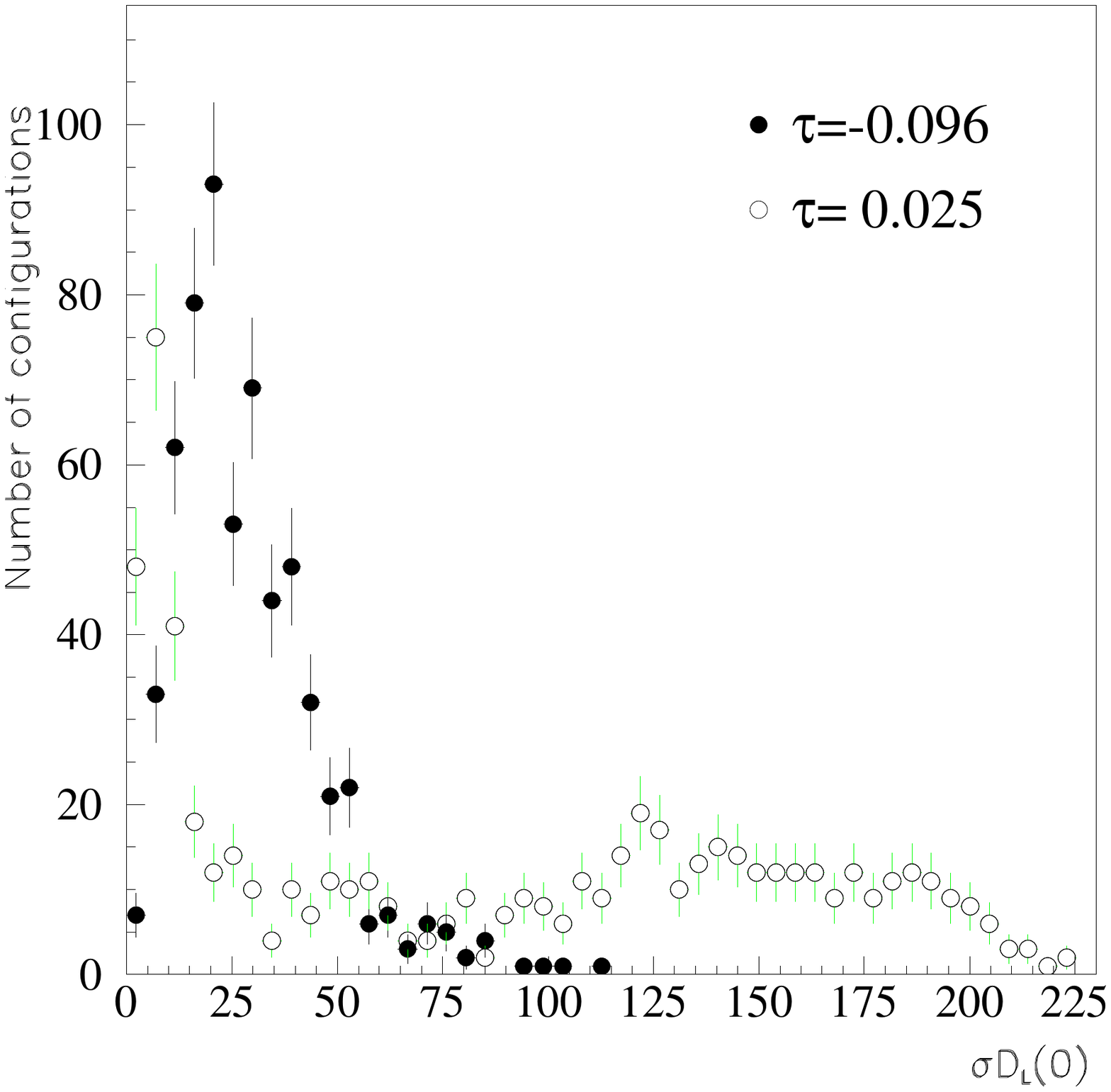}\includegraphics[width=7cm,clip]{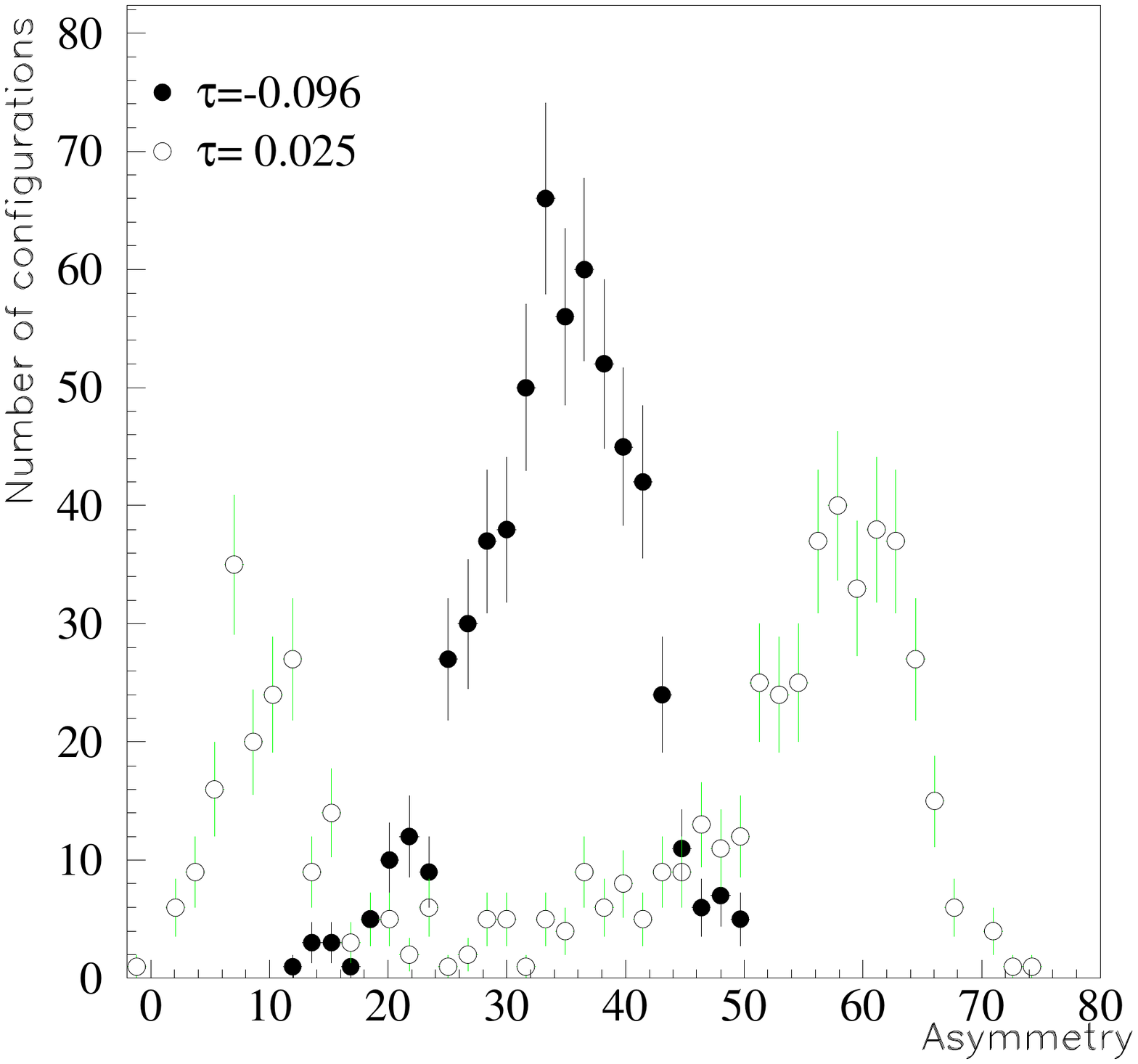}
\vspace*{-21mm}
\caption{Distributions in the zero-momentum longitudinal
propagator (left panel) and the asymmetry (right panel) at $T<T_c$ 
(filled circles) and $T>T_c$ (empty circles).}
\label{fig:DL0_asym_dist}       
\end{figure}

It is convenient to begin the analysis of the critical behavior
of the longitudinal $D_L(p)$ and transverse $D_T(p)$ propagators
by considering the temperature dependence 
of the distributions of gauge-field
configurations in $D_L(0)$. 

First, we observe that the distribution in 
$D_L(0)$ depends strongly on the temperature when $T\sim T_c$ (see Fig.~\ref{fig:DL0_asym_dist}). 
This behavior gets even more complicated due to finite-volume 
effects, which are considered to be substantial at $T\sim T_c$.

From the plots in Fig.~\ref{fig:DL0_asym_dist} it is clearly seen that 
not only the average value but also
 other quantities characterizing 
the distribution of confugurations in $D_L(0)$ should 
be thorougly analyzed in order to gain an insight
to the critical behavior of the propagators and 
the asymmetry $\aasymt$.
To characterize the propagator distribution,
it is instructive to consider its correlation with the 
Polyakov loop, whose temperature and volume dependence is 
well understood \cite{Gattringer:2010ms,Gattringer:2010ug}.
That is, we use the asymmetry as the reference. 

\begin{figure}[h]
\vspace*{-14mm}
\hspace*{-5mm}\includegraphics[width=7cm,clip]{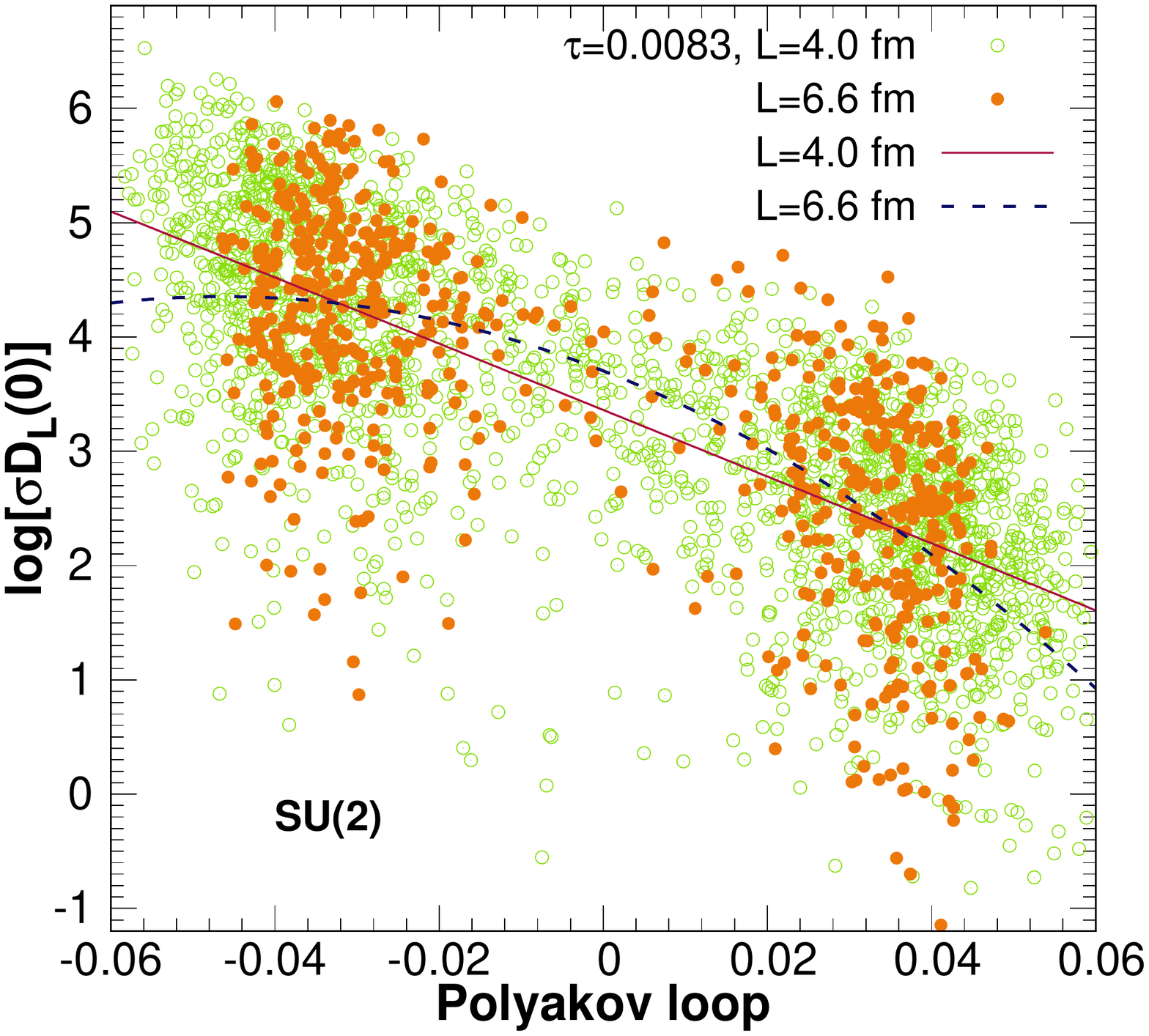}\hspace*{-4mm}\includegraphics[width=7cm,clip]{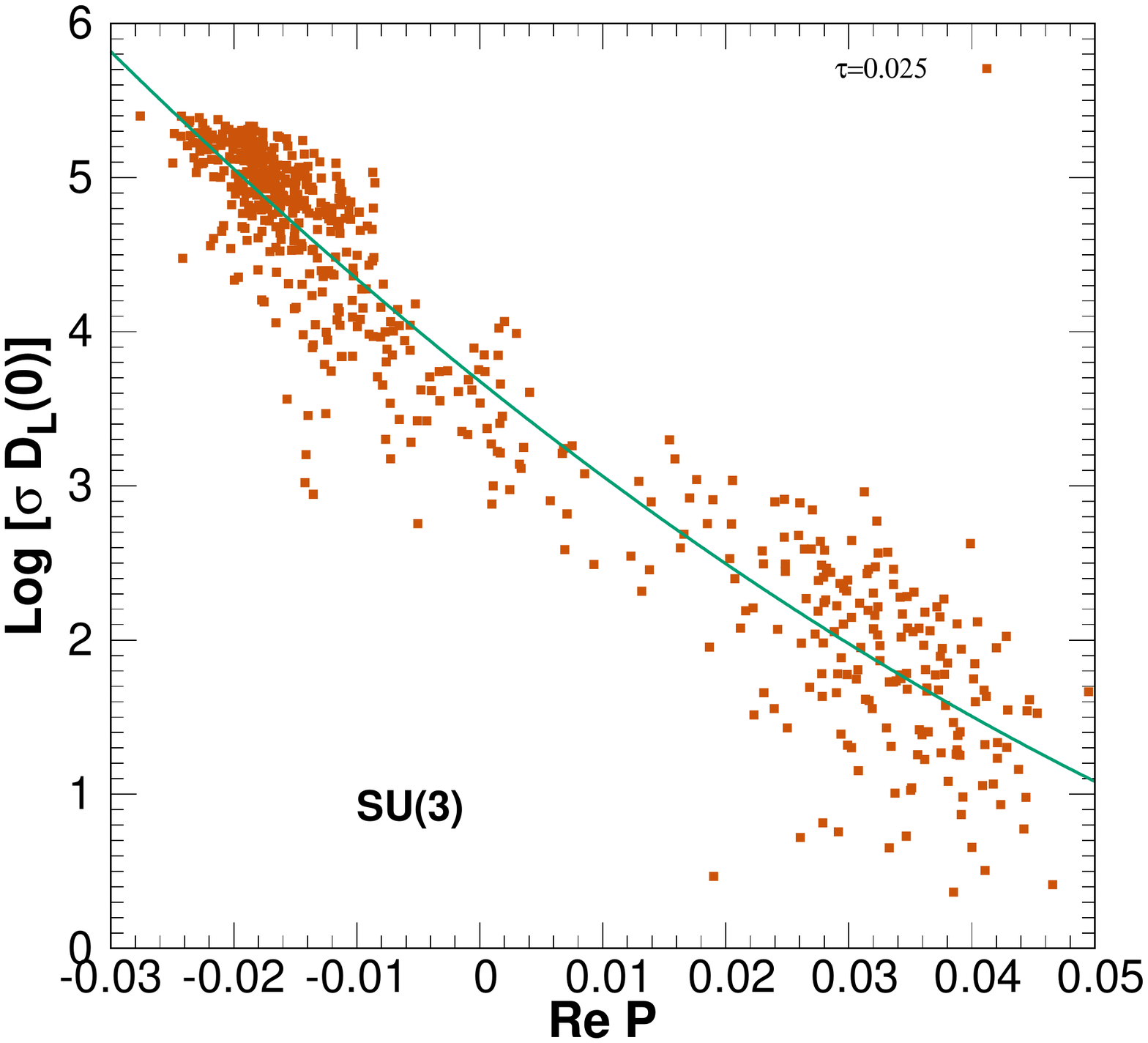}
\vspace*{-21mm}
\caption{Scatter plots illustrating correlation between the 
zero-momentum longitudinal propagator and the Polyakov loop. 
On the left panel data are shown at two different volumes and the same 
temperature. Shown lines are the regression curves based on the linear model (\ref{eq:DL_regression}).
}
\label{fig:DL0_ploo_corr}       
\end{figure}

Thus we consider the conditional distribution ${\cal F}\left(\mathbf{D}|{\cal P}\right)$ of $\mathbf{D}=\log\big[\sigma D_{L,T}(0)]$ at a particular value of the Polyakov loop ${\cal P}$ and perform regression analysis based on the linear regession model
\beq\label{eq:DL_regression}
 \langle \mathbf{D}(\tau)\rangle_{\cal P} \simeq {\cal D}_0^{\!(L,T)}(\tau) 
+ {\cal D}_1^{\!(L,T)}(\tau)  \operatorname{Re}{\cal P}(\tau)
+ {\cal D}_2^{\!(L,T)}(\tau) \big(\operatorname{Re}{\cal P}(\tau)\big)^2\;,
\eeq
where $\langle \mathbf{D}(\tau)\rangle_{\cal P}$ is the conditional average
of the logarithm of the respective normalized propagator at a given value of the Polyakov loop ${\cal P}$. The results for the longitudinal propagator are shown in Fig.~\ref{fig:DL0_ploo_corr}.

\begin{figure}[h]
\vspace*{-11mm}
\hspace*{-4mm}\includegraphics[width=7cm,clip]{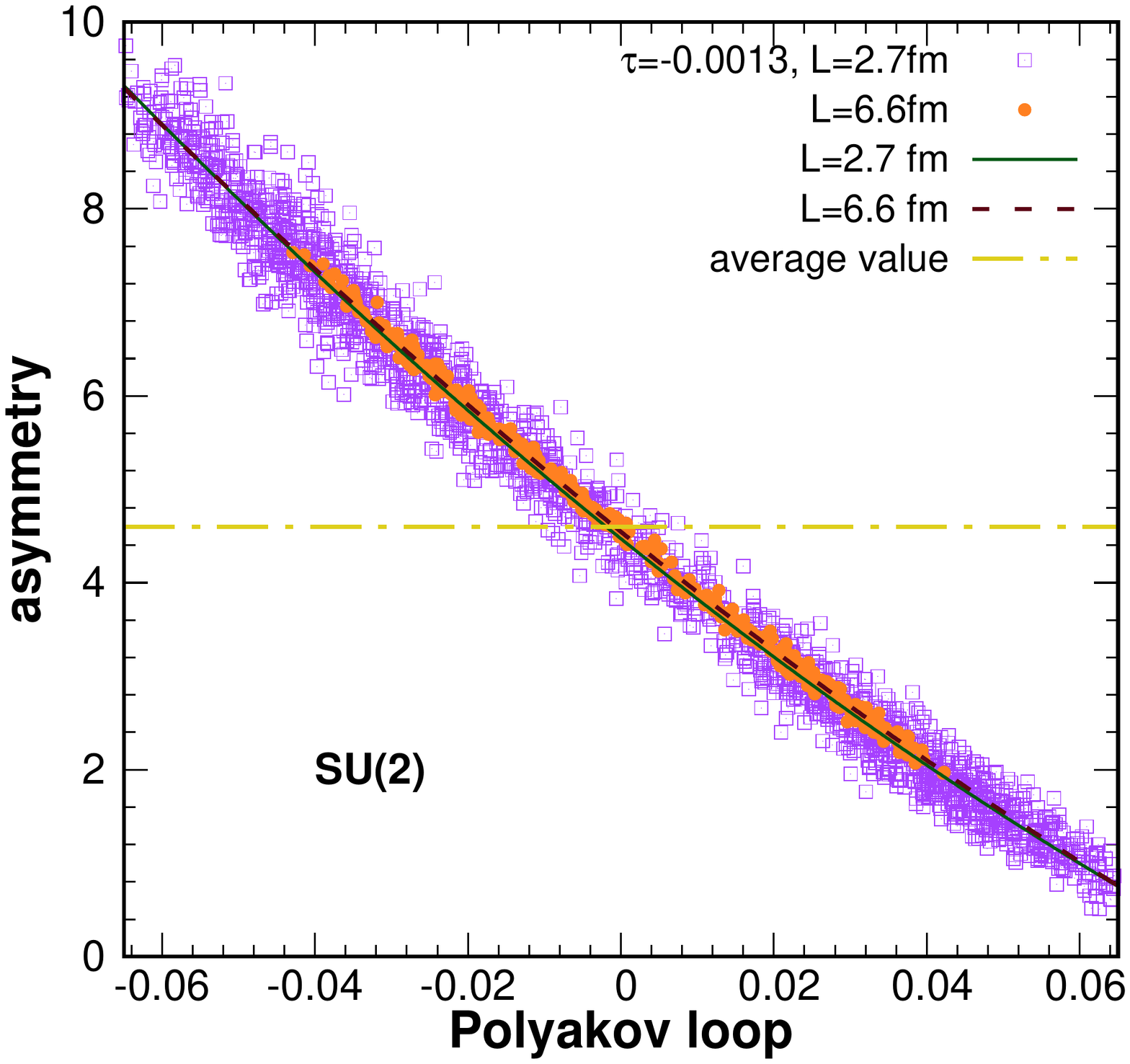}\hspace*{-4mm}\includegraphics[width=7cm,clip]{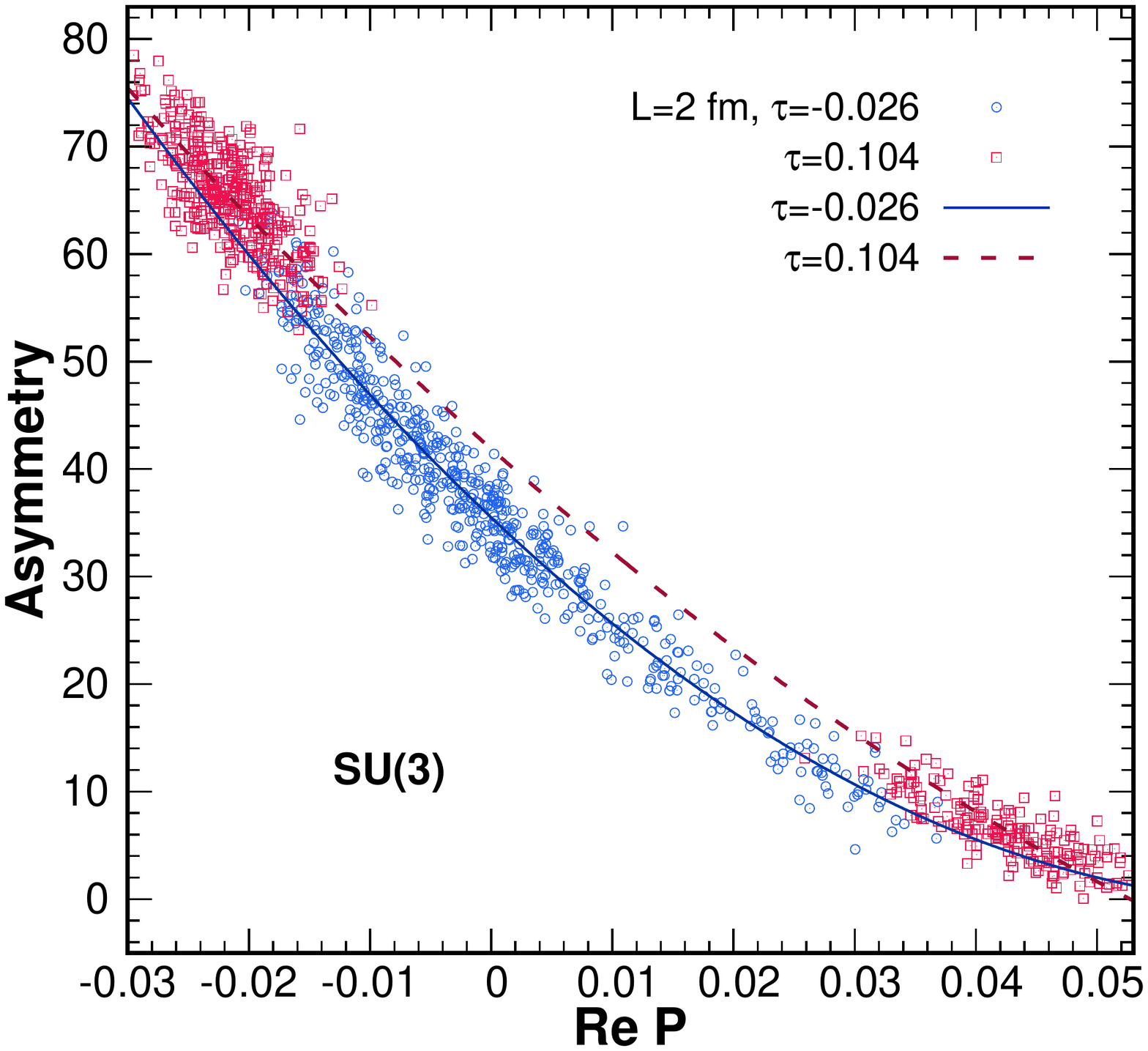}
\vspace*{-21mm}
\caption{Scatter plots illustrating correlation between the asymmetry and the Polyakov loop. Data are shown at two different lattice sizes and the same temperature in the SU(2) case (left panel) and at two different temperatures and the same lattice size in the SU(3) case (right panel).
Shown lines are the regression curves based on the linear model (\ref{eq:asym_regressed_fit}). 
 }
\label{fig:asym_pl_corr_FVE}       % Give a unique label
\end{figure}

A similar procedure is performed for the asymmetry, 
\beq
\label{eq:asym_regressed_fit}
 \langle {\cal A}\rangle_{\cal P} 
\simeq \aasymt_0(\tau) + \aasymt_1(\tau) \operatorname{Re}{\cal P}(\tau) + \aasymt_2(\tau) \big(\operatorname{Re}{\cal P(\tau)}\big)^2 \,,
\eeq
where 
the parameters $\aasymt_0$, $\aasymt_1$, and $\aasymt_2$ 
are also determined from the fit to data. The results are shown in Fig.~\ref{fig:asym_pl_corr_FVE}.

\section{Finite-volume effects and properties of the correlations}
\label{sec:FVE}

\begin{figure}[h]
\vspace*{-14mm}
\hspace*{-4mm}\includegraphics[width=7cm,clip]{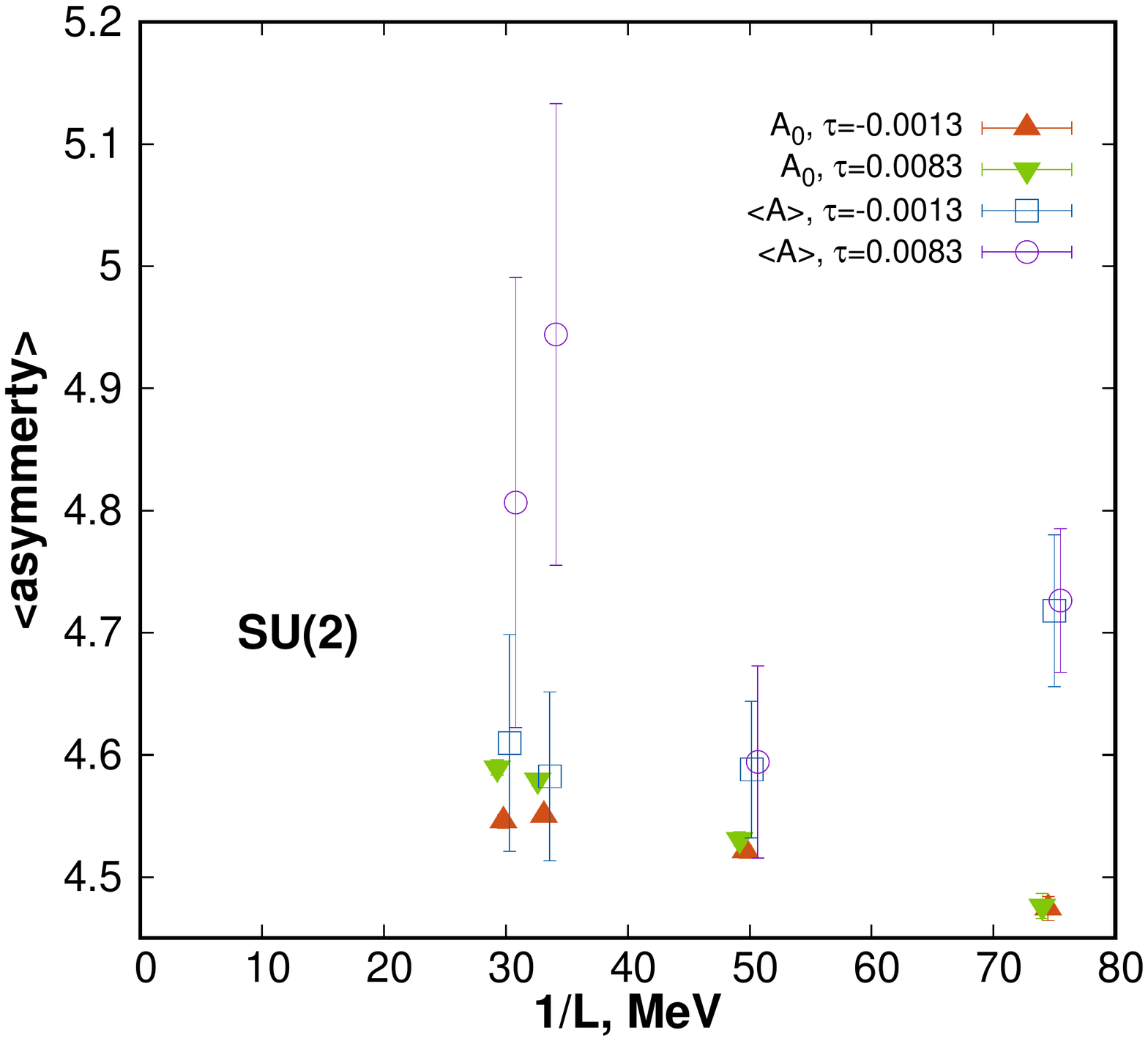}\hspace*{-4mm}\includegraphics[width=7cm,clip]{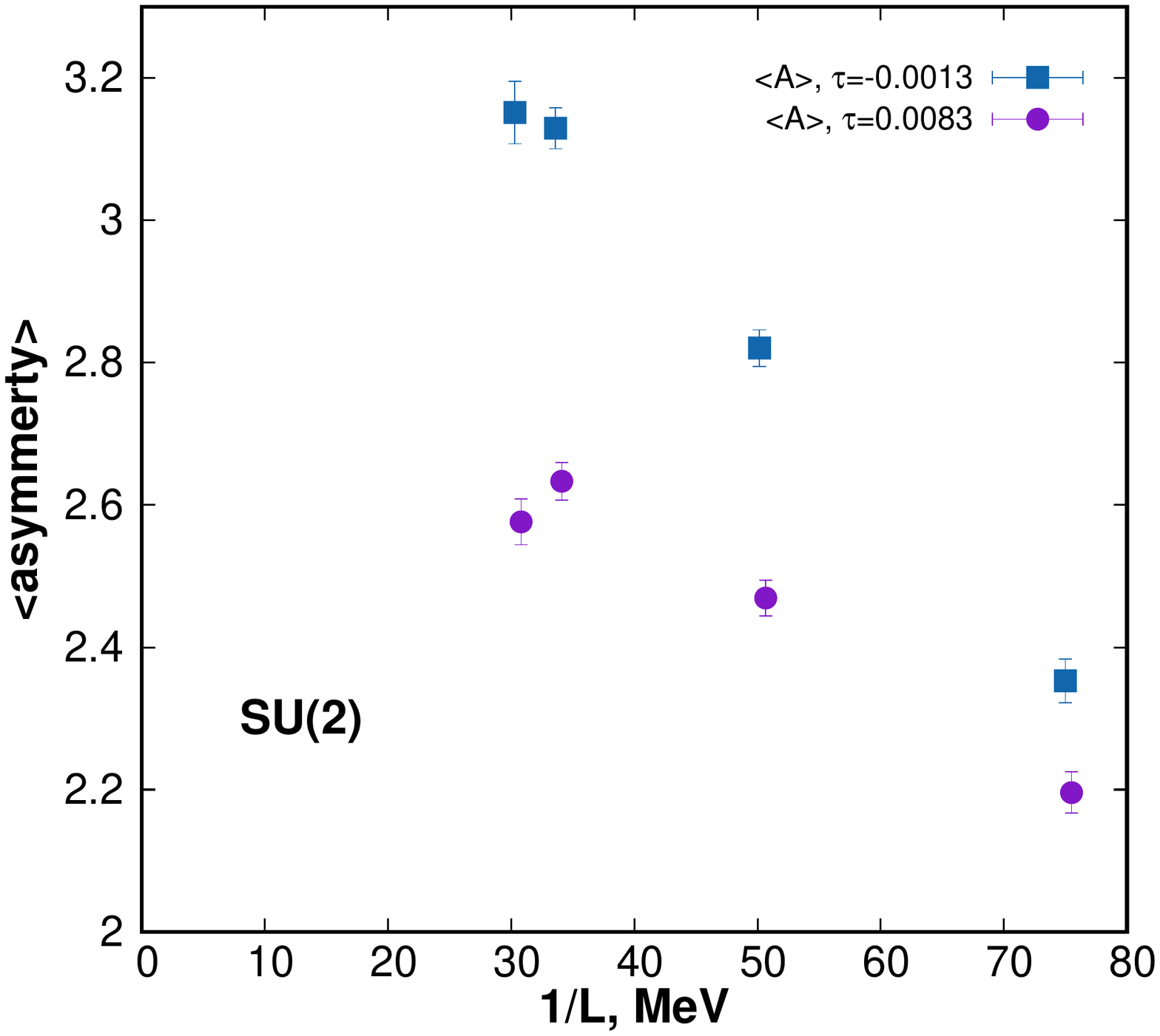}
\vspace*{-21mm}
\caption{Finite-volume effects for the asymmetry, $\aasymt_0$ $\langle \aasymt \rangle$ is the sample average. Left panel: both center sectors are taken into consideration. Right panel: results for the sector ${\cal P}>0$. }
\label{fig:FVE}       
\end{figure}

The right-hand side of the formula (\ref{eq:asym_regressed_fit})
is useful for an evaluation of the average value of the asymmetry 
in the infinite-volume limit
at the temprature $\tau$, where ${\cal P}(\tau)$ is the respective 
infinite-volume value of the Polyakov loop.
Firstly, the sample average value of the asymmetry
in the infinite-volume limit
coincides with the conditional average at ${\cal P}(\tau)$
because the variance of the Polyakov-loop distribution
tends to zero as $L\to\infty$.
Secondly, finite-volume effects for 
the predicted value (\ref{eq:asym_regressed_fit})
are much less than the sample average
as is shown in Fig.~\ref{fig:FVE}.
Infinite-volume extrapolation of the results shown on the right panel
must coincide with those on the left panel. Thus a comparison of panels
illustrates huge finite-volume effects due to an exclusion of one center sector,
which is common practice at $0<\tau<\!\!\!1$.
The left panel illustrates that the regression analysis
gives much better precision for the finite-volume 
expectation values of the asymmetry 
than the conventional use of the sample average.
It is also seen that an inclusion of both center sectors
results not only in substantial decrease 
of the finite-volume effects, but also 
in some decrease of precision.

\begin{figure}[h]
\vspace*{-14mm}
\hspace*{-4mm}\includegraphics[width=7cm,clip]{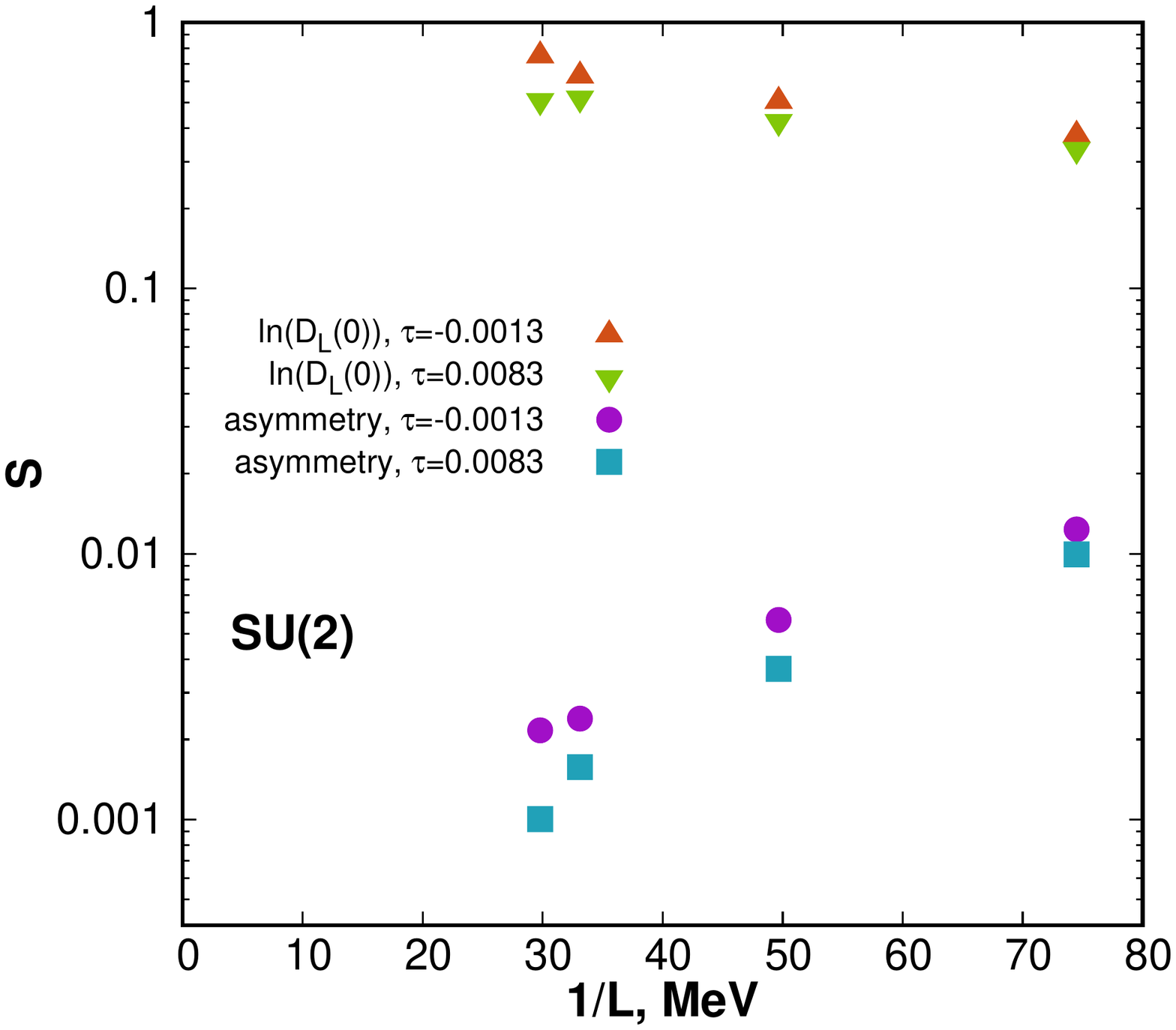}\hspace*{-4mm}\includegraphics[width=7cm,clip]{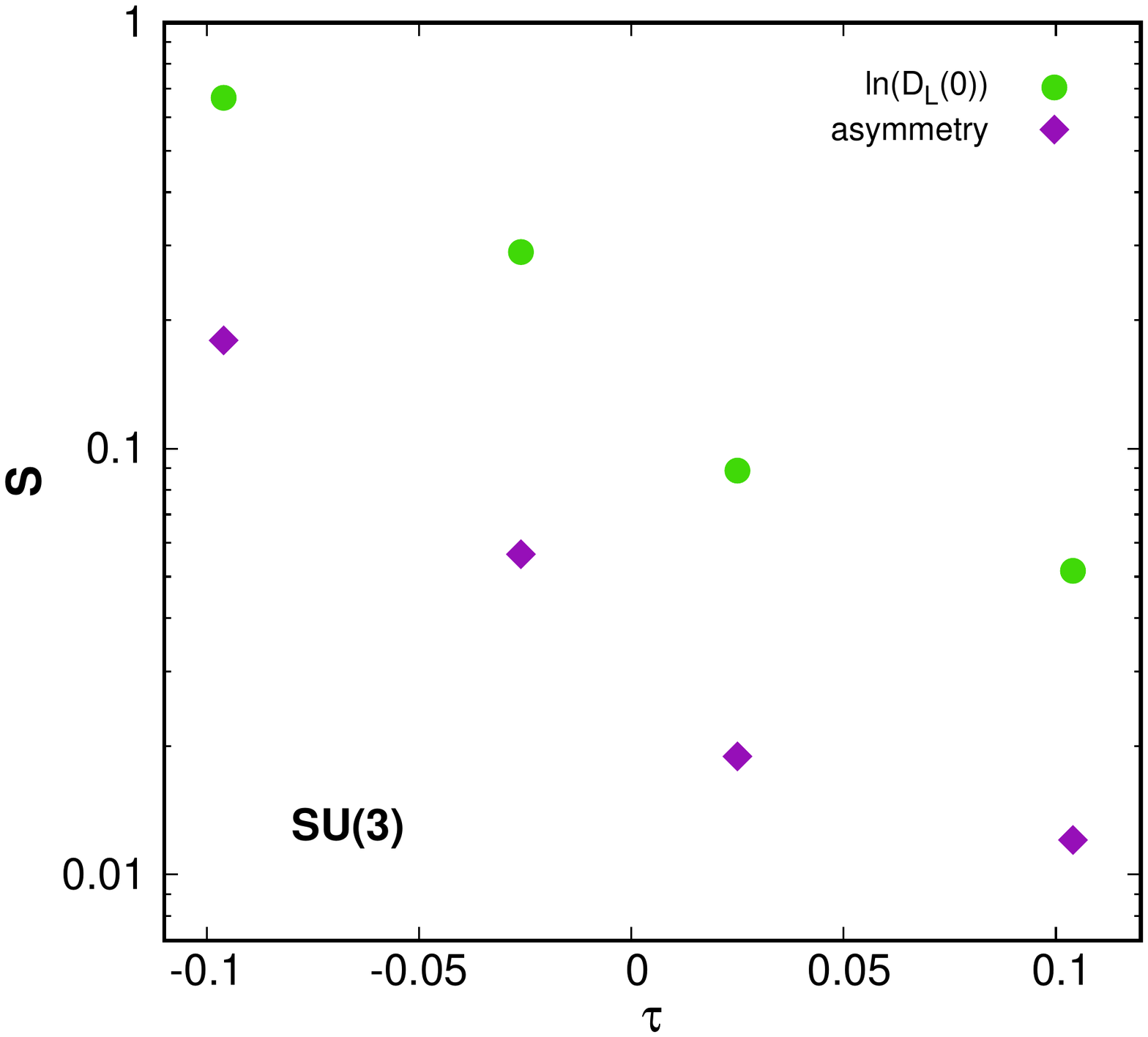}
\vspace*{-21mm}
\caption{Fraction of variance unexplained: the dependence on lattice size
is shown in the SU(2) case (left panel), the dependence on the temperature 
-- for the SU(3) case (right panel) both for $\aasymt$ and $\mathbf{D}$ }
\label{fig:Variance_Unexpl}       
\end{figure}

It is seen on the left panel of Fig.~\ref{fig:asym_pl_corr_FVE} that
its correlation with ${\cal P}$ becomes more pronounced with 
an increase of the volume.
To study the dependence of the correlation on the volume,
we use the Fraction of Variance Unexplained (FVU) designated by $S$
as a measure of the degree of correlation.
It represents the ratio of the variance of the residuals 
$ e_A(n) = {\cal A}_n - \aasymt_0 - \aasymt_1 \operatorname{Re}{\cal P}_n - \aasymt_2 (\operatorname{Re}{\cal P}_n)^2 $
and the sample variance $M_2(\aasymt)$,
%It is given by the formulas
\beq\label{eq:Variance_Unexpl}
S={ Q \over M_2 (\aasymt)}, \quad \mbox{where} \quad 
Q = \sum_{n=1}^{N_{data}} e^2_A(n)\,, \quad  
M_2(\aasymt)= \sum_{n=1}^{N_{data}} |{\cal A}_n - \langle {\cal A} \rangle|^2\,.
\eeq
In the general case, $0\leq S\leq 1$; $S=0$ implies that the asymmetry is nothing but a function of the Polyakov loop; $S=1$ means that they are independent of each other. The results are shown in Fig.~\ref{fig:Variance_Unexpl}.
A similar estimate of $S$ is performed also for the propagator.

We see that the FVU for the asymmetry tends to zero as $L\to\infty$,
whereas for the propagator it remains on the order of unity.
The temperature dependence of FVU is explained by taking all
center sectors into account at $T>T_c$ providing a wide range of 
Polyakov-loop variation.
We take all center sectors into consideration because in a finite volume 
transitions between them are not negligible even at $T>T_c$
when $\tau<\!\!\!<1$.

Our data give some evidence that, in the infinite-volume limit, 
the zero-momentum longitudinal propagator 
is not a proper function of the Polyakov loop,
it remains a random variable correlated with the Polyakov loop,
whereas the asymmetry becomes a proper 
function of the Polyakov loop. 

Nevertheless, our conclusion of the critical behavior of the propagator
presenrted in Refs.\cite{Bornyakov:2016geh,Bornyakov:2018mmf,Bornyakov:2021pls}
holds true because it is based on the assumption of the smooth dependence
of the conditional average $\langle \mathbf{D}(\tau)\rangle_{\cal P}$
on ${\cal P}$, no assumptions on its variance are made. 
Therewith, the regression curve can be extracted from the data
only with a limited precision, therefore,
the concept of smoothness should be formulated for 
the corridor of errors rather than for the proper function.

In this context, smoothness of the regression 
function  $f({\cal P})\equiv \langle \mathbf{D}(\tau)\rangle_{\cal P}$
at ${\cal P}={\cal P}_s$ means that the value
$f({\cal P}_s)$ as well as the derivative 
$\ds {df({\cal P})\over d{\cal P}}\Big|_{{\cal P}={\cal P}_s}$
evaluated over the range ${\cal P}<{\cal P}_s$
coincide within statistical error with those
evaluated over the range ${\cal P}>{\cal P}_s$.
It should be emphasized that our conclusions 
on the critical behavior are based on 
smothness at ${\cal P}_s=0$.

Our data fulfil this criterion. 
For example, if we take ${\cal P}_s=0$ and the 
SU(3) data for $\tau=-0.096$ so that the data come close to 
${\cal P}_s$, then 
we can use the fit function 
\beq
 \langle {\cal A}\rangle_{\cal P} 
\simeq \aasymt_0 + \aasymt_1 \operatorname{Re}{\cal P} 
\eeq
both at ${\cal P}<{\cal P}_s$ and at ${\cal P}>{\cal P}_s$; 
to put it differently, the third term in formula (\ref{eq:asym_regressed_fit})
can be omitted because ${\cal P}$ varies over a small range.
In so doing, we arrive at 
\bea
\aasymt_0=33.91(85), \quad \aasymt_1=-966(16) && \quad {\cal P}<0  \\ \nonumber 
\aasymt_0=33.90(64), \quad \aasymt_1=-962(91) && \quad {\cal P}>0\ .   \nonumber
\eea
However, the assumption of smooth dependence should be 
tested on a larger statistics, for which a reliable extrapolation 
to the point ${\cal P}_s=0$ is possible 
both from the domain ${\cal P}<0$ and ${\cal P}>0$ 
when $\tau>0$. 

Smooth dependence of 
the conditional averages $\langle\aasymt\rangle_{\cal P}$ and 
$\langle\mathbf{D}\rangle_{\cal P}$ on ${\cal P}$ at ${\cal P}=0$
implies that these quantities and, therefore, the infinte-volume 
limits of the respective sample averages considered as the functions of $T$ 
have the same singularity/nonanalyticity at $T=T_c$
that the Polyakov loop.

\section{Propagators at nonzero momenta}
\label{sec:glpr_pneq0}

\begin{figure}[h]
\vspace*{-11mm}
\hspace*{-4mm}\includegraphics[width=7cm,clip]{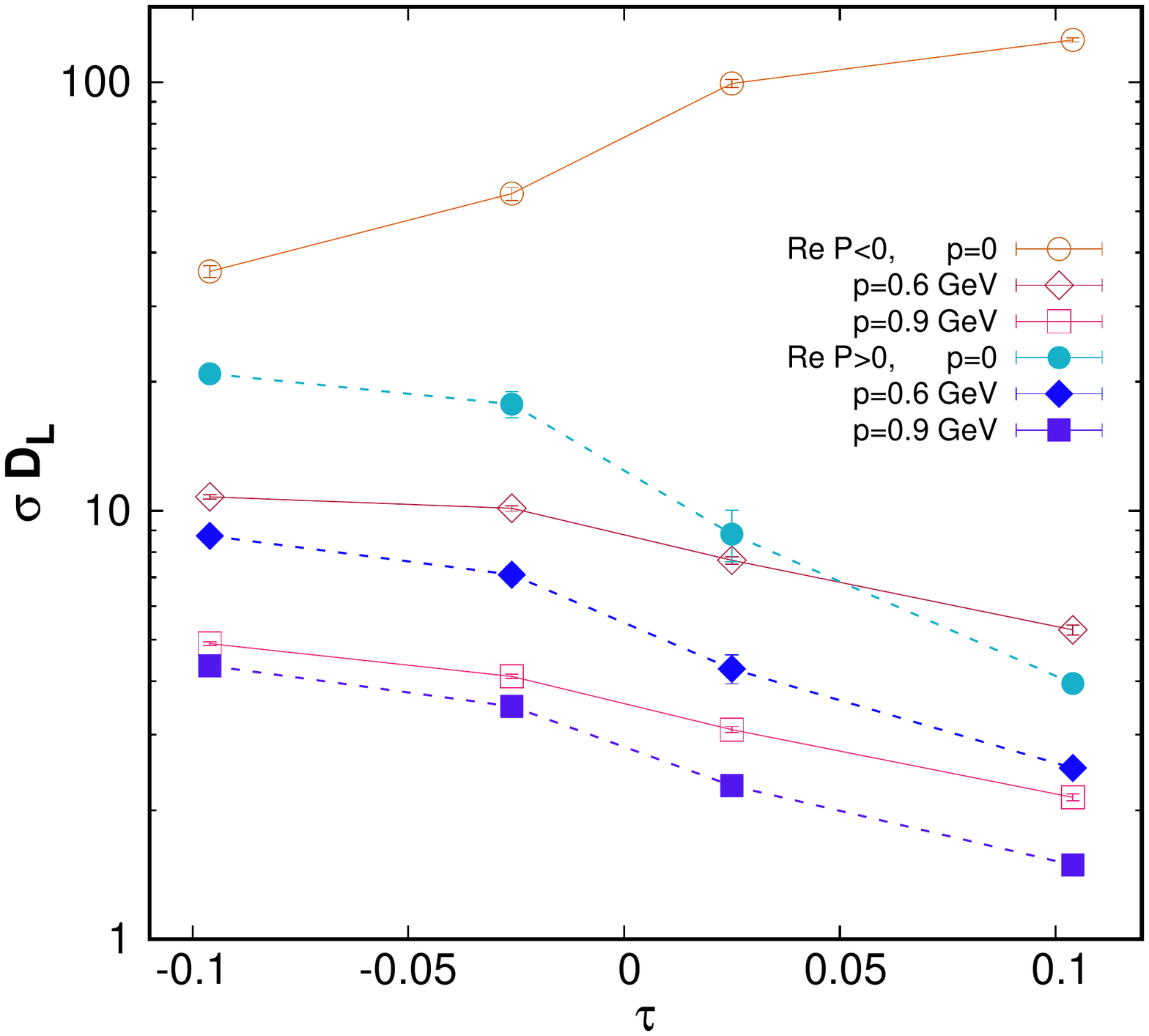}\hspace*{-4mm}\includegraphics[width=7cm,clip]{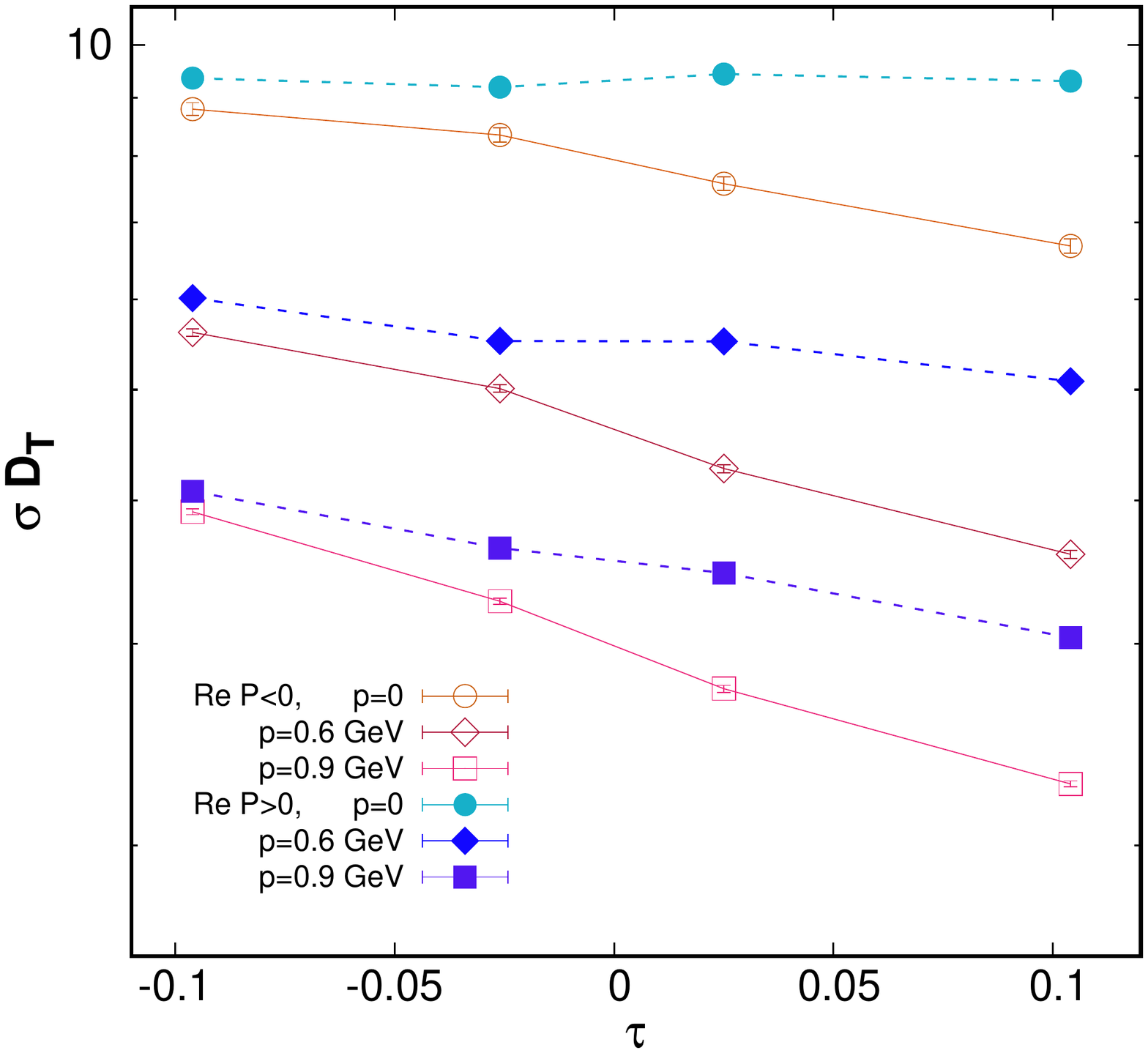}
\vspace*{-21mm}
\caption{Temperature dependence of the longitudinal
(left panel) and transverse (right panel) gluon propagators at 
various values of momenta in the SU(3) case. Lines are shown just to guide the eye, also notice the logarithmic scale on the ordinate axis.}
\label{fig:glpr_vs_T}       
\end{figure}

The mentioned test on smoothness can also be motivated by our observation that
the temperature dependence of the longitudinal propagator 
at nonzero momenta differs significantly from that at zero momentum,
which is shown in Fig.~\ref{fig:glpr_vs_T}.
It is clearly seen that the temperature dependence 
of the zero-momentum propagator is in a good agreement 
with that found in Refs.~\cite{Oliveira:2014uga,Silva:2016onh}. 

However, we find that the zero-momentum longitudinal 
gluon propagator decreases with the temperature
in the sector $\mathrm{Re}{\cal P}>0$
and rapidly increases in the sectors $\mathrm{Re}{\cal P}<0$,
whereas longitudinal gluon propagator at $p\sim 0.5\div 1$~GeV 
decreases with temperature in all Polyakov-loop sectors\footnote{Temperature dependence of $D_L(p)$ at $p\sim 0.5\div 1$~GeV 
is poorly seen on the plots in Refs.~\cite{Oliveira:2014uga,Silva:2016onh}:
one can only conclude that it does not increase.}. 
Thus a sharp peak of the longitudinal gluon propagator 
in the sectors $\mathrm{Re}{\cal P}<0$ appears in the deep infrared
at $T\sim T_c$. Physical interpretation of such peak may be as follows:
a formation of a large bubble
of the phase with $\mathrm{Re}{\cal P}<0$ in gluon matter at $T>T_c$
in the phase with $\mathrm{Re}{\cal P}>0$
gives rise to huge fluctuations of chromoelectric fields
which should emerge when the size of such bubble exceeds 
the scale determined by the width of the above-mentioned peak. 

Temperature dependence of the transverse propagator 
at nonzero momenta is shown in the right panel of 
Fig.~\ref{fig:glpr_vs_T}. It is interesting to remark that
it substantially decreases with temperature in the sector $\mathrm{Re}{\cal P}$.
However, this observation calls for further investigation. 

\section{Conclusions}
\label{conclu}

We have studied numerically the asymmetry and the 
longitudinal gluon propagator in the Landau-gauge 
SU(2) and SU(3) gluodynamics close to criticality.
Our findings can be summarized as follows:
 
\begin{itemize}
 \item The correlations between $\aasymt$ and ${\cal P}$
between  $\mathbf{D}$ and ${\cal P}$ indicate that the 
conditional averages $\langle\aasymt\rangle_{\cal P}$ and 
$\langle\mathbf{D}\rangle_{\cal P}$ are smooth functions of 
${\cal P}$ at ${\cal P}=0$.

\item In the infinite-volume limit, the asymmetry is completely determined by the 
Polyakov loop, whereas the propagators are only partially determined 
(that is, the gluon propagator at a given value of ${\cal P}$ represents
a random variable even in the infinite-volume limit. )

\item The use of regression analysis substantially reduces finite-volume effects.

\item Finite-volume effects in different center sectors partially 
cancel each other.

\item The temperature dependence of the longitudinal gluon propagator
at $p<1$~GeV and $T\sim T_c$ in the Polyakov-loop sectors 
with $\mathrm{Re}{\cal P}<0$ is qualitatively different at $p=0$
and $p=p_{min}\sim 600$~MeV.

\end{itemize}

 \vspace*{5mm}
{\bf Acknowledgments.} Computer simulations were performed on the IHEP 
(Protvino)  Central  Linux  Cluster and ITEP(Moscow) Linux Cluster.
This work was supported in part by 
the Russian Foundation for Basic Research, grant no.20-02-00737~A.

\bibliographystyle{woc}
\bibliography{ref_crit_behavior}

\end{document}